\documentclass[letterpaper,11pt,twoside]{article}
\usepackage{amsmath,amssymb,amsthm}
\usepackage{epic,eepic}

\newcommand{\xd}{\mathrm{d}}
\newcommand{\tens}{\otimes}
\newcommand{\defeq}{:=}
\newcommand{\Z}{\mathbb{Z}}
\newcommand{\C}{\mathbb{C}}

\DeclareMathOperator{\cou}{\epsilon}
\DeclareMathOperator{\cop}{\Delta}
\newcommand{\one}{\mathbf{1}}
\DeclareMathOperator{\id}{id}
\newtheorem{prop}{Proposition}
\newtheorem{thm}[prop]{Theorem}
\newtheorem{cor}[prop]{Corollary}
\newtheorem{lem}[prop]{Lemma}
\newtheorem{dfn}[prop]{Definition}
\newcommand{\Rop}{R}
\newcommand{\Qop}{Q}

\newcommand{\Gcompl}{G}
\newcommand{\Gfeyn}{G_F}
\newcommand{\Gconn}{G_c}

\newcommand{\compl}{\rho}
\newcommand{\conn}{\sigma}

\newcommand{\SV}{\mathsf{S}(V)}

\newcommand{\Dopnew}{\Omega}
\newcommand{\lp}{l}
\newcommand{\vertex}{v}
\newcommand{\Top}{T}
\newcommand{\fct}{\nu}
\newcommand{\edge}{e}

\allowdisplaybreaks

\title{\textbf{Generating loop graphs via Hopf algebra in quantum
    field theory}}
\author{\^Angela Mestre\footnote{email: mestre@matmor.unam.mx},
 Robert Oeckl\footnote{email: robert@matmor.unam.mx}\\ \\
Instituto de Matem\'aticas, UNAM, Campus Morelia,\\
C.~P.~58190, Morelia, Michoac\'an, Mexico}
\date{UNAM-IM-MOR-2006-1\\
18 July 2006\\ 5 October 2006 (v2)}

\begin{document}

\maketitle

\begin{abstract}
We use the Hopf algebra structure of the time-ordered algebra of field
operators to generate all connected weighted Feynman graphs in a
recursive and efficient manner. The algebraic representation of the
graphs is such that they can be evaluated directly as contributions to
the connected n-point functions. The recursion proceeds by loop order
and vertex number.

\end{abstract}

The combinatorics of perturbative quantum field theory is
traditionally dealt with via functional methods and
generating functions. However, it is possible to use a more intrinsic
algebraic approach instead, rooted directly at the level of $n$-point
functions and pioneered in the 1960's, see \cite{Rue:statmech}.

More recently, it was realized that the Hopf algebra structure of the
algebra of field operators (with the normal or with the time-ordered
product) can be fruitfully exploited. In particular, using the Hopf
algebra and its cohomology it was shown (among other things) how
different products of the algebra of field operators are related by
Drinfeld twists and how interactions correspond to 2-cocycles
\cite{BFFO:twist}.

Relations between different types of $n$-point functions and the
associated combinatorics of Feynman graphs via the Hopf algebra
structure of the time-ordered algebra of field operators was
established in \cite{MeOe:npoint}. More precisely, the relations
between complete and connected $n$-point functions on the one hand and
between connected and 1-particle irreducible $n$-point functions on
the other hand were described in this way. At the center of that work
stands an algorithm to recursively generate all tree graphs and their
values as Feynman graphs. The underlying structure in this is an
algebraic representation of graphs in terms of certain generalized
monomials in field operators.

In the present paper we extend this algorithm to recursively generate
all connected graphs using this algebraic representation. The
recursion proceeds by loop number (and by vertex number). The special
case of vanishing loop number precisely recovers the algorithm of
\cite{MeOe:npoint}. Crucially, and as in the special case of tree
graphs,
the correct weights of graphs are obtained so as to allow for their
direct evaluation in terms of the Feynman graph expansion of the
connected $n$-point function of a quantum field theory.
Note, however, that no type of renormalization procedure is taken into
account. In this sense the computed $n$-point functions may
be considered as bare ones.

As in the previous work \cite{MeOe:npoint} all results apply to bosonic
as well as fermionic fields and the algorithm is amenable to direct
implementation and should allow efficient calculations.

Section~\ref{sec:basics} reviews basics about certain graphs and their
symmetries, $n$-point functions, Feynman graphs, the algebraic
representation of graphs and the Hopf algebra structure of the
time-ordered field operator algebra. Section~\ref{sec:loop} contains
the main result with the algorithmic construction of connected graphs
and its proof.
Section~\ref{sec:recrel} presents an alternative recursive algorithm
to construct connected graphs that can be applied directly on the
level of $n$-point functions.
Section~\ref{sec:discuss} offers some discussion,
especially concerning the efficient algorithmic implementation and the
inclusion of fermions.
The appendix lists all connected graphs without external edges and
with up to three internal edges together with their weight factors.

\section{Basic concepts and definitions}
\label{sec:basics}

The basic setup in this paper is substantially similar to that of
\cite{MeOe:npoint}. Hence, the present section has substantial overlap
with Section~II and a part of Section~IV of that paper. Nevertheless,
there are important differences, most importantly a more extensive treatment
of abstract graphs and their symmetries.

\subsection{Graphs}
\label{sec:graphs}

We introduce certain kinds of graphs and elementary properties of
them. The graphs will be later interpreted as Feynman graphs. Here we
are only interested in them as abstract graphs.

\begin{dfn}
A \emph{graph} is a finite collection of \emph{vertices} and
\emph{edges}, such that any
end of an edge may be connected to a vertex.
Edges that are connected to vertices at both ends are
called \emph{internal}, while edges with at least one free end are
called \emph{external}. Internal edges with both ends connected to the
same vertex are also called \emph{self-loops}.
The \emph{valence} of a vertex is the number of ends of edges
connected to the vertex. The \emph{loop number} of a graph is its
number of cycles. A graph is \emph{connected} if it is connected as a
topological space.
\end{dfn}

We recall the well known relation between vertex number, edge number
and loop number of connected graphs:

\begin{lem}
\label{lem:loopnumber}
Consider a connected graph with at least one vertex.
Let $\vertex,\edge,\lp$ be its number of vertices,
internal edges and loops, respectively. Then,
\begin{equation}
\label{eq:loopnumber}
\lp=\edge-\vertex+1\,.
\end{equation}
\end{lem}  

\begin{dfn}
A \emph{labeled} graph is a graph whose free ends of external edges
are labeled with labels from a label set. Labels on different ends of
edges are required to be distinct.
\end{dfn}

In the following we shall consider only such labeled graphs, i.e.,
from now on \emph{graph} means \emph{labeled graph}. The label set is
fixed from the outset and will later be identified with an appropriate
set of field operator labels.

\begin{dfn}
\label{def:order}
A graph is said to be \emph{vertex ordered} if its vertices are
ordered. That is, the vertices are numbered $1,\dots,n$ where $n$ is
the total number of vertices.
A graph is said to be \emph{edge ordered} if the ends of its internal
edges are ordered. That is, the ends of internal edges are numbered
$1,\dots,2 n$ where $n$ is the total number of internal edges (each
edge having two ends).
A graph is called \emph{ordered} if it is both
vertex ordered and edge ordered.
\end{dfn}

\begin{dfn}
\label{def:symfac}
Consider an ordered graph $\gamma$. A \emph{symmetry} of $\gamma$ is a
permutation of the numbering of the vertices and of the endpoints of
the internal edges that yields combinatorially the same ordered (and labeled)
graph. The number of symmetries, i.e., the order of the group of
permutations leaving the graph invariant, is called the \emph{symmetry
factor} of the graph. It will be denoted by $S^\gamma$.
\end{dfn}
 
Since the symmetry factor is the same for any ordering of the vertices
and ends of internal edges of a graph, the concept is well defined for
unordered graphs as well.

\begin{dfn}
\label{def:vertexsymfac}
Consider a vertex ordered graph $\gamma$. A \emph{vertex symmetry} of
$\gamma$ is a permutation of the numbering of its vertices, which
yields combinatorially the same vertex ordered (and labeled) graph. The
order of the group of vertex symmetries is called the \emph{vertex
  symmetry factor} of the graph. It will be denoted by
$S^\gamma_{\text{vertex}}$.
\end{dfn}

\begin{dfn}
\label{def:edgesymfac}
Consider an ordered graph $\gamma$.  An \emph{edge symmetry} of
$\gamma$ is a permutation of the numbering of the ends of its
internal edges that yields combinatorially the same ordered (and
labeled) graph while the order of the vertices is held fixed. The
order of the group of edge symmetries is called the \emph{edge
  symmetry factor} of the graph. It will be denoted by
$S^\gamma_{\text{edge}}$.
\end{dfn}

Clearly, the concepts of  vertex and edge symmetry factors also make
sense for unordered graphs as the vertex and edge symmetry factors are
the same for any ordering of the vertices and of the ends of the
internal edges of a graph, respectively.

\begin{lem}
\label{lem:sym}
Let $\gamma$ denote an ordered graph.
The orders of the associated symmetry groups satisfy
$S^{\gamma}=
S^{\gamma}_{\text{vertex}}\cdot
S^{\gamma}_{\text{edge}}$.
\end{lem}
\begin{proof}
Denote the group of symmetries, vertex symmetries and edge symmetries
of $\gamma$ by $G^\gamma$, $G^\gamma_\text{vertex}$ and
$G^\gamma_\text{edge}$ respectively. Note that an edge symmetry is
merely a particular type of symmetry. Hence,
the group $G^\gamma_\text{edge}$ may be seen as a subgroup of
$G^\gamma$ via an injective group
homomorphism $G^\gamma_\text{edge}\to G^\gamma$. Furthermore, a
symmetry defines a vertex symmetry by forgetting its action on the
numbering of ends of edges. On the other hand, any vertex symmetry
can be augmented to a symmetry. Hence, there is a surjective group
homomorphism $G^\gamma\to G^\gamma_\text{vertex}$.
It is easy to see that the group homomorphisms form an exact sequence of
groups
\[
 0\to G^\gamma_\text{edge}\to G^\gamma\to G^\gamma_\text{vertex}\to 0 .
\]
Hence, as the groups are finite their orders
satisfy $S^{\gamma}=
S^{\gamma}_{\text{vertex}}\cdot
S^{\gamma}_{\text{edge}}$.
\end{proof}

\begin{lem}
\label{lem:edgesymfac}
Consider a connected graph $\gamma$.
Let $v$ be its number of
vertices. For each vertex $1\le i\le v$ let $p_i$ be the number of
self-loops connected to it.
Let $t$ be the number of pairs of
vertices which are directly connected through at least one edge.
For each pair $1\le j\le t$ of such vertices let $q_j$ be the number of
edges connecting it. Then, the edge
symmetry factor of $\gamma$ is given by
$S^\gamma_{\text{edge}}=(\prod_{i=1}^\vertex\, 2^{p_i}\cdot
p_i!)(\prod_{j=1}^t\,q_j!)$.
\end{lem}

The proof is straightforward combinatorics.

\subsection{$n$-point functions and Feynman graphs}
\label{sec:feyn}

The physical content of a quantum field theory is usually
extracted from its $n$-point functions. In perturbation theory, these
are computed as sums of values of Feynman graphs. We briefly
review here the essentials. More details can be found in any standard
text book on quantum field theory such as \cite{ItZu:qft}.

We denote by
$\Gcompl^{(n)}(x_1,\dots,x_n)$ the \emph{complete} 
$n$-point function. This is the vacuum expectation value of
the time-ordered product of $n$ field operators, i.e.,
\[
 \Gcompl^{(n)}(x_1,\dots,x_n)
 =\langle 0 | T \phi(x_1)\cdots \phi(x_n) | 0\rangle .
\]
The notation we use here suggests a scalar field theory on Minkowski
spacetime. In general there would be internal field indices as well
and possibly other modifications (other spacetime etc.).
The real nature of the fields is completely irrelevant
for our treatment as long as the standard perturbative treatment applies.
Therefore, we shall continue
with our present notation for simplicity. Hence, we denote field operators
generically by $\phi(x)$, where $x$ is from a label set (here
suggestive of points in Minkowski space).
Furthermore, we shall assume all fields to
be bosonic. The fermionic case is also straightforward, but includes
extra factors, see Section~\ref{sec:discuss}.

Let $V$ be the complex vector space of linear combinations of field
operators $\phi(x)$. The algebra generated by the field operators with
the \emph{time-ordered} product is commutative and can be identified
with the \emph{symmetric algebra} $\SV$ over $V$. More precisely,
$\SV=\bigoplus_{k=0}^\infty V^k$, where $V^k$ is the space of 
linear combinations of monomials of degree $k$ in the field operators
and $V^0$ is the one-dimensional vector space spanned by the identity
element $\one$.
We may now express ensembles of $n$-point functions as functions
$\SV\to\C$. In particular, we may set
\[
\compl(\phi(x_1)\cdots\phi(x_n)) \defeq \Gcompl^{(n)}(x_1,\dots,x_n) .
\]

In perturbation theory, the $n$-point functions can be computed
as a sum over values of Feynman graphs. For the complete $n$-point
functions we may write
\begin{equation}
 \Gcompl^{(n)}(x_1,\dots,x_n)
 =\sum_{\gamma\in\Gamma^n} w_\gamma \gamma(x_1,\dots,x_n) .
\label{eq:npoint}
\end{equation}
Here $\Gamma_n$ is the set of Feynman graphs. These are graphs
$\gamma$ in the sense of Section~\ref{sec:graphs} with $n$
external edges labeled by field operator labels
$x_1,\dots,x_n$.\footnote{Note that usually Feynman graphs involve
lines of different type depending on particle species. In our treatment
lines correspond to sums over all particle species. The information
about which particle species can interact resides completely in the
vertex functions.}
Indeed, from here onward we
fix the label set to be the label set of the field operators.
The value of a
graph $\gamma$
labeled by $x_1,\dots,x_n$ is denoted above by
$\gamma(x_1,\dots,x_n)$. The set $\Gamma_n$ may be taken to be
precisely the set of all graphs with $n$ external legs (up to
topological equivalence). The
weight factor $w_\gamma$ is precisely the inverse of the symmetry
factor $S^\gamma$ of a graph in the sense of
Definition~\ref{def:symfac}.

We should emphasize that the discussion here applies to bare
$n$-point functions. Renormalization is outside the scope of the
present paper.

The type of $n$-point functions we shall be interested in in the
following are the \emph{connected} ones, denoted $\Gconn^{(n)}$. These
may be defined in the same way as (\ref{eq:npoint}), but with the
restriction that only connected graphs are considered. We define
$\sigma:\SV\to \C$ via
\[
 \conn(\phi(x_1)\cdots\phi(x_n)) \defeq \Gconn^{(n)}(x_1,\dots,x_n).
\]

We now turn to the calculation of the value of a Feynman graph.
The Feynman
propagator $\Gfeyn(x,y)$ is the value of the graph that consists of an
edge only, its two ends labeled by $x$ and $y$
respectively.
The value of a graph that consists of a vertex with external edges
labeled by $x_1,\dots,x_k$ is given by the vertex
function $F(x_1,\dots,x_k)$. Note that we can encode the ensemble of
vertex functions in a way analogous to $n$-point functions as a
function $\nu:\SV\to\C$ via
\begin{equation}
 \nu(\phi(x_1)\cdots\phi(x_n))\defeq F(x_1,\dots,x_n) .
\label{eq:vfunc}
\end{equation}

For more general graphs we also need the inverse Feynman propagator,
$\Gfeyn^{-1}$, determined by the 
equation
\begin{equation}
\int \xd y\, \Gfeyn(x,y)\Gfeyn^{-1}(y,z)=\delta(x,z) .
\label{eq:invfey}
\end{equation}
The value of a general graph may then be computed as follows:
Associate a label with each internal edge and form the product over
a vertex function associated with each vertex and an inverse Feynman
propagator associated to each internal edge. Finally, integrate over
all possible assignments of internal labels.

\subsection{Algebraic representation of graphs}
\label{sec:grep}

We introduce an algebraic representation of graphs based on the
time-ordered operator algebra $\SV$ and allowing straightforward
evaluation of graphs in the above sense. More precisely, we associate
a given graph with $v$ vertices with a certain element in
$\SV^{\tens\vertex}$, the $v$-fold tensor product of $\SV$.

Each
vertex of the graph corresponds to one tensor factor. A product
$\phi(x_1)\cdots\phi(x_n)$ in a
given tensor factor corresponds to external edges of the associated
vertex whose endpoints are labeled by $x_1,\dots,x_n$.
To represent internal edges, we define the formal elements
$\Rop_{i,j}\in\SV^{\tens \vertex}$ with
$1\le i\le j\le \vertex$ using the inverse Feynman propagator
(\ref{eq:invfey}).\footnote{$\Rop_{i,j}$ is
formal insofar as it  really lives in
a completion of the tensor product $\SV^{\tens\vertex}$. However,
this fact is largely irrelevant for our purposes.}
For $i\neq j$ the definition is
\begin{equation}\label{eq:Rij}
\Rop_{i,j} := \int
\xd x\,\xd y\,G_F^{-1}(x,y)\,(\one^{\tens i-1}\tens \phi(x)
\tens\one^{\tens j-i-1}
\tens\phi(y)\tens\one^{\tens v-j}) ,
\end{equation}
with the field operators $\phi(x)$ and $\phi(y)$
inserted at the $i^{\mbox{\tiny{th}}}$ and
$j^{\mbox{\tiny{th}}}$ positions, respectively.
For $i=j$ the definition is
\begin{equation}\label{eq:Rii}
\Rop_{i,i} := \int
\xd x\,\xd y\,G_F^{-1}(x,y)\,(\one^{\tens i-1}\tens \phi(x)
\phi(y)\tens\one^{\tens v-i}) ,
\end{equation}
The element $\Rop_{i,j}\in\SV^{\tens\vertex}$ corresponds to one
internal edge connecting
the vertices which occupy the positions $i$ and $j$. In particular,
the element $\Rop_{ i,i}\in\SV^{\tens\vertex}$ for $1\le i\le\vertex$
is interpreted as an internal edge connecting the
$i^{\mbox{\tiny{th}}}$ vertex to itself. That is, it corresponds to a
self-loop.

\begin{figure}
\setlength{\unitlength}{0.00083333in}
\begingroup\makeatletter\ifx\SetFigFont\undefined%
\gdef\SetFigFont#1#2#3#4#5{%
  \reset@font\fontsize{#1}{#2pt}%
  \fontfamily{#3}\fontseries{#4}\fontshape{#5}%
  \selectfont}%
\fi\endgroup%
{\renewcommand{\dashlinestretch}{30}
\begin{picture}(5412,2204)(0,-10)
\texture{44555555 55aaaaaa aa555555 55aaaaaa aa555555 55aaaaaa aa555555 55aaaaaa 
	aa555555 55aaaaaa aa555555 55aaaaaa aa555555 55aaaaaa aa555555 55aaaaaa 
	aa555555 55aaaaaa aa555555 55aaaaaa aa555555 55aaaaaa aa555555 55aaaaaa 
	aa555555 55aaaaaa aa555555 55aaaaaa aa555555 55aaaaaa aa555555 55aaaaaa }
\path(4974,1950)(4824,1500)
\path(4974,1950)(4824,1500)
\path(4074,1950)(4239,1500)
\path(4074,1950)(4239,1500)
\path(4074,1050)(4224,1500)
\path(4074,1050)(4224,1500)
\path(4974,1050)(4824,1500)
\path(4974,1050)(4824,1500)
\put(2511,1438){\makebox(0,0)[lb]{{\SetFigFont{12}{14.4}{\rmdefault}{\mddefault}{\updefault}$x_2$}}}
\put(1321,1443){\makebox(0,0)[lb]{{\SetFigFont{12}{14.4}{\rmdefault}{\mddefault}{\updefault}$x_1$}}}
\path(2471,1506)(1561,1506)
\path(4209,1485)(4779,1485)
\path(4209,1485)(4779,1485)
\put(4254,1500){\shade\ellipse{240}{240}}
\put(4254,1500){\ellipse{240}{240}}
\put(4809,1500){\shade\ellipse{240}{240}}
\put(4809,1500){\ellipse{240}{240}}
\put(2011,1676){\ellipse{160}{346}}
\put(2018,1508){\shade\ellipse{240}{240}}
\put(2018,1508){\ellipse{240}{240}}
\put(128,1478){\shade\ellipse{240}{240}}
\put(128,1478){\ellipse{240}{240}}
\put(91,76){\makebox(0,0)[lb]{{\SetFigFont{12}{14.4}{\rmdefault}{\mddefault}{\updefault}$\one$}}}
\put(1231,73){\makebox(0,0)[lb]{{\SetFigFont{12}{14.4}{\rmdefault}{\mddefault}{\updefault}$\Rop_{1,1}\cdot(\phi(x_1)\phi(x_2))$}}}
\put(3256,58){\makebox(0,0)[lb]{{\SetFigFont{12}{14.4}{\rmdefault}{\mddefault}{\updefault}$\Rop_{1,2}\cdot(\phi(x_1)\phi(x_2)\otimes\phi(x_3)\phi(x_4))$}}}
\put(4951,2033){\makebox(0,0)[lb]{{\SetFigFont{12}{14.4}{\rmdefault}{\mddefault}{\updefault}$x_3$}}}
\put(4944,848){\makebox(0,0)[lb]{{\SetFigFont{12}{14.4}{\rmdefault}{\mddefault}{\updefault}$x_4$}}}
\put(3946,2033){\makebox(0,0)[lb]{{\SetFigFont{12}{14.4}{\rmdefault}{\mddefault}{\updefault}$x_1$}}}
\put(3946,833){\makebox(0,0)[lb]{{\SetFigFont{12}{14.4}{\rmdefault}{\mddefault}{\updefault}$x_2$}}}
\end{picture}
}
\caption{Examples of the algebraic representation of graphs in terms
of elements of $\SV^{\tens \vertex}$.}
\label{fig:exacor}
\end{figure}
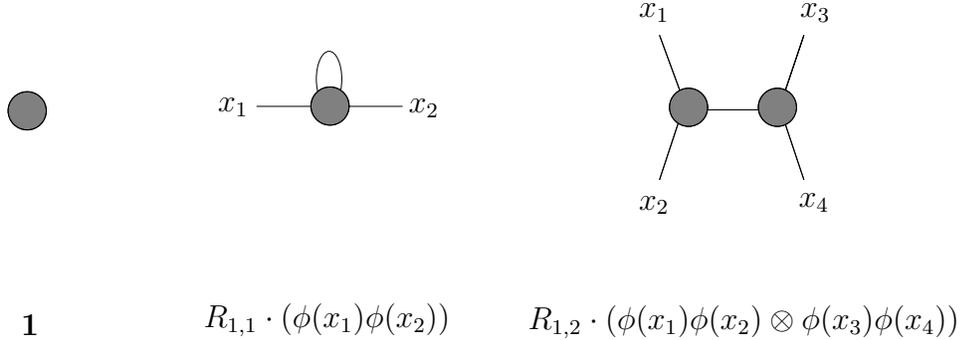

Combining several internal edges (which can be self-loops) and their
products with external edges by multiplying the respective
expressions in $\SV^{\tens \vertex}$ allows to build arbitrary graphs
with $\vertex$ vertices. Figure~\ref{fig:exacor} shows some
examples. It is then obvious that applying the vertex functions $\nu$
defined by (\ref{eq:vfunc}) to each tensor factor yields precisely the
value of the respective graph as a Feynman graph. Thus, the graphs
we just discussed are exactly those
that are to enter in the $\vertex$-vertex  contribution to an
$n$-point function.

The ordering of the tensor factors of $\SV^{\tens \vertex}$
induces an ordering of the vertices of the graphs in the sense of
Definition \ref{def:order}. However, when applying
$\fct^{\tens \vertex}$ the ordering is ``forgotten''. Indeed, it is not
relevant for the interpretation of graphs as Feynman graphs, but
only plays a role at
the level of their algebraic representation here. 
In the following, we will encounter elements of $\SV^{\tens \vertex}$
that are linear combinations of expressions corresponding to
graphs. In this context, we call the scalar multiplying the expression
for a given graph the \emph{weight} of the graph.
Clearly, if we are interested in unordered graphs, the weight of such
a graph is the sum of the weights of all vertex ordered graphs that
correspond to it upon forgetting the vertex order.

\subsection{The field operator algebra as a Hopf algebra}

A crucial ingredient of our setting is the fact that the algebra $\SV$
of time-ordered field operators is not only an algebra, but
a \emph{Hopf algebra}. That is, $\SV$ carries a coproduct $\cop:\SV\to
\SV\tens \SV$ and a counit $\cou:\SV\to\C$ that are compatible with
its algebra structure and unit. ($\SV$ also carries an antipode map,
but this will not be used in the following.)
We refer the reader to \cite{Swe:hopfalg}
for a classical treatment of Hopf algebras and to \cite{Lan:algebra}
for the Hopf algebra structure of the symmetric algebra. The
significance of this Hopf
algebra structure for quantum field theory was developed in
\cite{BFFO:twist}. (Note,
however, that the product taken there is the normal product and not
the time-ordered one.)

At this point we will only mention the
explicit form of the coproduct on $\SV$. On monomials this takes the
form
\begin{equation}
\cop (\phi(x_1)\cdots \phi(x_n))=
 \sum_{I_1\cup I_2 = \{\phi(x_1),\dots,\phi(x_n)\}} T(I_1)\tens T(I_2),
\label{eq:coproduct}
\end{equation}
and is extended to all of $\SV$ by linearity.
Here the sum runs over partitions of the set of field operators
$\{\phi(x_1),\dots,\phi(x_n)\}$ into two sets $I_1$ and $I_2$. $T$
denotes the time-ordered product of the field operators in the
corresponding partition. The coproduct may be extended (by suitable
composition with itself) to a map (on monomials and extended by
linearity),
\begin{equation}
\cop^k (\phi(x_1)\cdots \phi(x_n))=
 \sum_{I_1\cup \cdots\cup I_{k+1} = \{\phi(x_1),\dots,\phi(x_n)\}}
 T(I_1)\tens \cdots \tens T(I_{k+1}) .
\label{eq:copk}
\end{equation}
The difference to the single coproduct is that the set of field
operators is now split into $k+1$ partitions. Note also that the
partitions are \emph{ordered}, i.e., the sets $I_1,\dots, I_{k+1}$ are
distinguishable. An important property of the coproduct is that it is
multiplicative, i.e., it is an algebra map with respect to the algebra
structure of $\SV$ (and the induced algebra structure on the tensor
product).
A more
extensive discussion of the Hopf algebra structure, adapted to the
present context can be found in \cite{MeOe:npoint}.

\section{Generating loop graphs}
\label{sec:loop}

\subsection{Statement of result}

The main result of this paper, which is the subject of the present
section, may be described as an efficient
algorithm that recursively generates all connected graphs $\gamma$.
The graphs are generated together with the correct weights $w_\gamma$,
explained in Section~\ref{sec:feyn}.
In particular, the recursion is such
that it may be organized in ascending loop order. Also, the graphs are
generated directly in the algebraic representation introduced in
Section~\ref{sec:grep}. This allows their direct evaluation as Feynman
graphs.

More precisely, we shall construct recursively a set of maps
$\Dopnew^{\lp,\vertex}:\SV\to\SV^{\tens\vertex}$ indexed
by integers $l$ and $v$ such that the following theorem holds.

\begin{thm}
\label{thm:main}
Fix integers $\lp, n\ge 0$, $\vertex\ge 1$ and operator labels
$x_1,\dots,x_n$. Then, $\Dopnew^{\lp,
  \vertex}(\phi(x_1)\cdots\phi(x_n))\in\SV^{\tens \vertex}$
corresponds to the weighted sum over all connected graphs with  $\lp$
loops, $\vertex$  vertices
and $n$ external edges whose endpoints are  labeled by
$x_1,\dots,x_n$,
each with weight being the inverse of its symmetry factor.
\end{thm}

This specializes for $l=0$ to Lemma~10 of \cite{MeOe:npoint}, with
$\Lambda^{\vertex-1}=\Dopnew^{0,\vertex}$.

We may conclude with the interpretation in terms of Feynman graphs
and $n$-point functions. Denote the $\lp$-loop and $\vertex$-vertex
contribution to the ensemble $\conn$ of connected $n$-point functions
by $\conn^{\lp,\vertex}$. In particular, the $\lp$-loop order
contribution $\conn^\lp$ to $\conn$ and $\conn$ itself are given by
\begin{equation*}
\conn^\lp=\sum_{\vertex=0}^\infty \conn^{\lp,\vertex},\qquad
\conn=\sum_{\lp=0}^\infty \conn^\lp .
\end{equation*}
There is only one contribution with zero vertex number. This is the
Feynman propagator contributing to the 2-point function. Hence
$\conn^{l,v}$ is zero if $v=0$ and $l\neq 0$, while $\conn^{0,0}$
is non-zero only on $V\tens V$ and coincides there with the Feynman
propagator. All non-zero vertex number contributions are captured by
the following corollary.

\begin{cor}
\label{cor:loopexpansion}
For $v\ge 1$:
\[
 \conn^{\lp,\vertex}=\fct^{\tens \vertex}\circ\Dopnew^{\lp,\vertex} .
\]
\end{cor}

Restricting to the $l=0$ (tree level) contribution recovers
Corollary~18 of \cite{MeOe:npoint}. Note, however, that the
contribution corresponding to the Feynman propagator was missing there
as well as in Theorem~5 of that paper.

The appendix lists all connected graphs without
external edges as weighted contributions to
$\Dopnew^{\lp,\vertex}(\one)$, for edge number $\edge=l+v-1\le 3$.

\subsection{Construction and proof}

The proof proceeds in a manner very analogous to the proof in
\cite{MeOe:npoint}. Indeed, each intermediate lemma in this section
specializes to a corresponding lemma in Section~IV of that paper when
restricted to the case $l=0$. We do not point this out explicitly in
the following, but refer the reader to that paper for comparison.

In order to construct $\Dopnew^{\lp,\vertex}$ we introduce certain
auxiliary maps.
Using the component-wise product in $\SV^{\tens\vertex}$, we may view
the elements
$\Rop_{i,j}$, defined by equations (\ref{eq:Rij}) and (\ref{eq:Rii}),
as operators on this
space by multiplication. In particular, these elements are used to
define the following maps:
\begin{itemize}
\item
$\Top_i:\SV^{\tens\vertex}\to\SV^{\tens\vertex}$, with $1\le i\le
  \vertex$, as the operator $\Rop_{i,i}$ together with the factor
  $1/2$:
\begin{equation}
\Top_i\defeq\frac{1}{2}\Rop_{i,i}\,.
\label{eq:T}
\end{equation}
\item
$\Qop_i:\SV^{\tens\vertex}\to \SV^{\tens{\vertex+1}}$, with $1\le i\le
  \vertex$, given by the composition of $\Rop_{i,i+1}$ with the
  coproduct applied to the $i^{\mbox{\tiny{th}}}$ component of
  $\SV^{\tens\vertex}$, i.e. $\cop_i\defeq
  \id^{\tens{i-1}}\tens\cop\tens\id^{\tens{\vertex-i}}:
  \SV^{\tens\vertex}\to\SV^{\tens{\vertex+1}}$,
  together with a factor of $1/2$:
\begin{equation}
\Qop_i \defeq \frac{1}{2} \Rop_{i,i+1}\circ\cop_i .
\label{eq:Q}
\end{equation} 
\end{itemize}

The map $\Top_i$ given by (\ref{eq:T}), endows the
$i^{\mbox{\tiny{th}}}$ vertex of a vertex ordered graph with a
self-loop together with a factor $1/2$. The latter is the inverse of
the edge symmetry factor of a single self-loop (see
Lemma~\ref{lem:edgesymfac}).
The action of the map $\Qop_i$
given by (\ref{eq:Q}) is less simple. Consider the coproduct $\cop_i$
applied to the
$i^{\mbox{\tiny{th}}}$ component of $\SV^{\tens \vertex}$.
Recalling the formula (\ref{eq:coproduct}), we
see that $\cop_i$ converts a graph with $\vertex$ vertices into a sum
over graphs with $\vertex+1$ vertices by \emph{splitting} the
$i^{\mbox{\tiny{th}}}$ vertex into two in all possible ways. That is,
the $i^{\mbox{\tiny{th}}}$ vertex is replaced by two vertices
(numbered $i$ and $i+1$) and the edges ending on it, (considered as
distinguishable)
are distributed between the two new vertices in all possible ways.
Note that the two new vertices are distinguished due to the ordering
of the tensor factors. Thus, to obtain the corresponding
operation for unordered graphs we need to divide by a factor of $2$.
This factor corresponds to the two different relative orderings of the
new vertices with which each unordered configuration occurs.
The only exception to this is the case when the split vertex has no
edges at all. No overcounting happens in this case.
The meaning of the map $\Qop_i$ given by (\ref{eq:Q})
becomes clear now in terms of graphs. Namely, it splits the
$i^{\mbox{\tiny{th}}}$ vertex into two and subsequently reconnects the
two new vertices with an edge.  Dividing by two compensates for the
double counting as described above if we are interested in unordered
graphs (assuming
the set of endings of edges of the split vertex is not empty). 

We remark that the maps $\Top_i$ increase both the loop and edge
numbers of a graph by one unit, leaving the vertex number invariant,
while the maps $\Qop_i$ increase both the edge and vertex numbers by
one unit, leaving the loop number invariant.

We use the maps $\Top_i$ and $\Qop_i$, given by equations
(\ref{eq:T}) and (\ref{eq:Q}), respectively, to define recursively
maps  $\Dopnew^{\lp,\vertex} :\SV\to\SV^{\tens\vertex}$ for
$\lp\ge 0$ and for $\vertex\ge 1$ as follows:
\begin{align}
\Dopnew^{0,1} & \defeq \id\,, \nonumber\\
\Dopnew^{\lp,\vertex} & \defeq
\frac{1}{\lp+\vertex-1}\left(\sum_{i=1}^{\vertex-1}\Qop_i \circ
\Dopnew^{\lp,\vertex-1}+\sum_{i=1}^{\vertex}
\Top_i\circ\Dopnew^{\lp-1,\vertex}\right)\,.
 \label{eq:recomega}
\end{align}

\begin{figure}
\begin{center}
\setlength{\unitlength}{0.00083333in}
\begingroup\makeatletter\ifx\SetFigFont\undefined%
\gdef\SetFigFont#1#2#3#4#5{%
  \reset@font\fontsize{#1}{#2pt}%
  \fontfamily{#3}\fontseries{#4}\fontshape{#5}%
  \selectfont}%
\fi\endgroup%
{\renewcommand{\dashlinestretch}{30}
\begin{picture}(1805,2484)(0,-10)
\path(970,2220)(754,1842)
\path(787.489,1961.073)(754.000,1842.000)(839.584,1931.305)
\path(967,2219)(1183,1841)
\path(1097.416,1930.305)(1183.000,1841.000)(1149.511,1960.073)
\path(668,1473)(452,1095)
\path(485.489,1214.073)(452.000,1095.000)(537.584,1184.305)
\path(665,1472)(881,1094)
\path(795.416,1183.305)(881.000,1094.000)(847.511,1213.073)
\path(1262,1473)(1046,1095)
\path(1079.489,1214.073)(1046.000,1095.000)(1131.584,1184.305)
\path(1259,1472)(1475,1094)
\path(1389.416,1183.305)(1475.000,1094.000)(1441.511,1213.073)
\path(971,726)(755,348)
\path(788.489,467.073)(755.000,348.000)(840.584,437.305)
\path(968,725)(1184,347)
\path(1098.416,436.305)(1184.000,347.000)(1150.511,466.073)
\path(371,723)(155,345)
\path(188.489,464.073)(155.000,345.000)(240.584,434.305)
\path(368,722)(584,344)
\path(498.416,433.305)(584.000,344.000)(550.511,463.073)
\path(1564,718)(1348,340)
\path(1381.489,459.073)(1348.000,340.000)(1433.584,429.305)
\path(1561,717)(1777,339)
\path(1691.416,428.305)(1777.000,339.000)(1743.511,458.073)
\put(610,1558){\makebox(0,0)[lb]{{\SetFigFont{12}{14.4}{\rmdefault}{\mddefault}{\updefault}$\Dopnew^ {0,2}$}}}
\put(1805,58){\makebox(0,0)[lb]{{\SetFigFont{12}{14.4}{\rmdefault}{\mddefault}{\updefault}$\Dopnew^ {3,1}$}}}
\put(905,2313){\makebox(0,0)[lb]{{\SetFigFont{12}{14.4}{\rmdefault}{\mddefault}{\updefault}$\Dopnew^ {0,1}$}}}
\put(1205,1558){\makebox(0,0)[lb]{{\SetFigFont{12}{14.4}{\rmdefault}{\mddefault}{\updefault}$\Dopnew^ {1,1}$}}}
\put(1505,803){\makebox(0,0)[lb]{{\SetFigFont{12}{14.4}{\rmdefault}{\mddefault}{\updefault}$\Dopnew^ {2,1}$}}}
\put(905,808){\makebox(0,0)[lb]{{\SetFigFont{12}{14.4}{\rmdefault}{\mddefault}{\updefault}$\Dopnew^ {1,2}$}}}
\put(1205,58){\makebox(0,0)[lb]{{\SetFigFont{12}{14.4}{\rmdefault}{\mddefault}{\updefault}$\Dopnew^ {2,2}$}}}
\put(605,58){\makebox(0,0)[lb]{{\SetFigFont{12}{14.4}{\rmdefault}{\mddefault}{\updefault}$\Dopnew^ {1,3}$}}}
\put(302,808){\makebox(0,0)[lb]{{\SetFigFont{12}{14.4}{\rmdefault}{\mddefault}{\updefault}$\Dopnew^ {0,3}$}}}
\put(0,58){\makebox(0,0)[lb]{{\SetFigFont{12}{14.4}{\rmdefault}{\mddefault}{\updefault}$\Dopnew^ {0,4}$}}}
\end{picture}
}
\end{center}
\caption{Recursive dependencies of the maps
$\Dopnew^{\lp,\vertex}$ up to order $\lp+\vertex\le 4$. The right
  directed arrows correspond to $\Top_i$ maps while the left directed
  ones correspond  to $\Qop_i$ maps.}
\label{fig:table}
\end{figure}
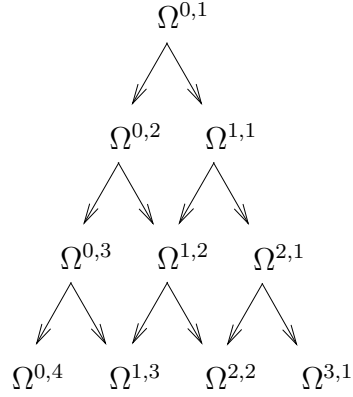

Note that in the recursion equation above the $\Top$ and $\Qop$-
summands do not appear when $\lp=0$ or when $\vertex=1$,
respectively. Figure \ref{fig:table} shows the recursive dependencies
of $\Dopnew^{\lp,\vertex}$ for different $\lp$, $\vertex$ with
$\lp+\vertex\le 4$.

We notice that $\Dopnew^{\lp,\vertex}$ satisfies the following
factorization property:
\begin{lem}
\label{lem:omegafac}
Fix integers $\lp,n, m\ge 0$, $\vertex\ge 1$ and operator labels
$x_1$, $\dots$, $x_n$, $y_1$, $\dots$,
$y_m$. Then,
$\Dopnew^{\lp,\vertex}$ satisfies the factorization property
\begin{multline}\label{eq:facomega2}
\Dopnew^{\lp,\vertex}(\phi(x_1)\cdots\phi(x_n)\phi(y_1)\cdots\phi(y_m))=\\
 \Dopnew^{\lp,\vertex}(\phi(x_1)\cdots\phi(x_n))\cdot
 \cop^{\vertex-1}(\phi(y_1)\cdots\phi(y_m)) .
\end{multline}
\end{lem}
\begin{proof}
This follows from the multiplicativity of the coproduct
and the recursive definition (\ref{eq:recomega}), noticing that each
time the vertex number increases by one, one coproduct is applied as
part of the operator $Q_i$.
\end{proof}

We now turn to the proof of Theorem~\ref{thm:main}. We begin with
weaker lemmas, increasing their strength stepwise until reaching the
desired result.

\begin{lem}
\label{lem:allgraphs}
Fix integers $\lp,n\ge 0$, $\vertex\ge 1$ as well as field
operator labels
$x_1,\dots,x_n$.
(a) $\Dopnew^{\lp,\vertex}(\phi(x_1)\cdots\phi(x_n))$
corresponds  to a weighted sum of  connected graphs with
$\lp$ loops, $\vertex$ vertices and $n$ external edges whose endpoints
are labeled by
$x_1,\dots,x_n$. (b) Any connected graph with  $\lp$
loops, $\vertex$
vertices and the given  external edges occurs
in $\Dopnew^{\lp,\vertex}(\phi(x_1)\cdots\phi(x_n))$ with some
positive weight.
\end{lem}
\begin{proof}
First, it is clear that $\Dopnew^{0,1}(\phi(x_1)\cdots\phi(x_n))$
corresponds to the connected graph with one single vertex with no
self-loops and the external edges whose endpoints are
labeled by $x_1,\dots,x_n$. Moreover,
$\Dopnew^{\lp,\vertex}(\phi(x_1)\cdots\phi(x_n))$ is generated from
this by sums
of multiple applications of the maps $\Top_i$ and  $\Qop_i$ with
scalar factors according to the recursion formula
(\ref{eq:recomega}). Both $\Top_i$ and $\Qop_i$ convert a
term corresponding to a connected graph to a sum over terms
corresponding to connected graphs. Thus,
$\Dopnew^{\lp,\vertex}(\phi(x_1)\cdots\phi(x_n))$ is a sum of terms
each of which
corresponds to a connected graph (with some weight). Second, the fact
that every graph contained in
$\Dopnew^{\lp,\vertex}(\phi(x_1)\cdots\phi(x_n))$ has $\lp$ loops and
$\vertex$ vertices follows from  noticing that the maps $\Top_i$
increase the loop number  by one unit, while the vertex number remains
fixed, and the maps $\Qop_i$ increase the vertex number by one unit,
leaving the loop number  unchanged. This concludes the proof of (a).

To prove (b) we proceed by induction on the internal edge number
$\edge=\lp+\vertex-1$ (recall Lemma~\ref{lem:loopnumber}). The result
is evidently valid for $\edge=0$, corresponding to $\lp=0$ and
$\vertex=1$. We assume the result holds for $\edge-1$. Let $\gamma$
denote a graph with $\lp$ loops and $\vertex$ vertices so that
$\lp+\vertex-1=\edge$. We show that it is generated by applying the
maps $\Top_i$ or $\Qop_i$ to graphs contained in
$\Dopnew^{\lp-1,\vertex}(\phi(x_1)\cdots\phi(x_n))$ or in
$\Dopnew^{\lp,\vertex-1}(\phi(x_1)\cdots\phi(x_n))$,
respectively. Since both $\Top_i$ and $\Qop_i$ produce graphs with
positive weight from graphs with  positive weight, the weight of a
graph $\gamma$  occurring in $\Dopnew^{\lp,\vertex}$, being given by a
sum over positive contributions according to formula
(\ref{eq:recomega}), is positive. Now, suppose the graph $\gamma$ has
at least one vertex with one or more self-loops. Let this vertex
occupy the $i^{\mbox{\tiny{th}}}$ position, for instance.   Shrinking
one of these self-loops yields a graph that corresponds
by assumption to a term in
$\Dopnew^{\lp-1, \vertex}(\phi(x_1)\cdots\phi(x_n))$ with some
positive weight so that applying the map $\Top_i$ to the vertex $i$
produces the graph $\gamma$ with (positive) weight. Thus, by  formula
(\ref{eq:recomega}) the  graph $\gamma$ occurs in
$\Dopnew^{\lp,\vertex}$. Finally, suppose the  graph $\gamma$ does not
contain vertices with self-loops. Choose an arbitrary internal
edge. Shrinking this edge and
fusing the vertices it connects yields a graph that corresponds
by assumption to a term in
$\Dopnew^{\lp, \vertex-1}(\phi(x_1)\cdots\phi(x_n))$. Say, the
fused vertex has position $i$. Applying $\Qop_i$ to this term will
yield a sum over terms one of which will correspond to the
original. By the recursive definition of
$\Dopnew^{\lp,\vertex}(\phi(x_1)\cdots\phi(x_n))$ it thus contains
this term with
positive weight. This completes the proof of (b).
\end{proof}

What remains in order to prove Theorem~\ref{thm:main} is to show
that the term corresponding to each graph  has weight given exactly by
the inverse of its symmetry factor. We start with a more
restricted result.
\begin{lem}
\label{lem:weightextedges}
Fix integers  $\lp\ge0$, $\vertex\ge 1$ and $n\ge \vertex$ as well
as field operator
labels $x_1,\dots,x_n$. Consider a connected graph $\gamma$ with
$\lp$ loops, $\vertex$ vertices, $n$ external
edges whose endpoints are labeled by $x_1,\dots,x_n$ and
the property
that each vertex has
at least one external edge ending on it. Then, the term in
$\Dopnew^{\lp, \vertex}(\phi(x_1)\cdots\phi(x_n))$ corresponding to
that graph has
weight given by the inverse of its symmetry factor $S^\gamma$.
\end{lem}
\begin{proof}
We proceed by induction on the number of internal edges
$\edge$. Clearly, the
statement is true for $\edge=0$ so that we assume it  holds for a
general number of internal edges $\edge-1$. Let $\gamma$ be a
connected graph with $\edge$ internal edges. Let $\lp$  be its loop
number and let $\vertex$ be
its vertex number. By Lemma~\ref{lem:allgraphs}, this graph  occurs in
$\Dopnew^{\lp, \vertex}(\phi(x_1)\cdots\phi(x_n))$ with positive
weight $\alpha$. We proceed to show that
$\alpha=1/S^\gamma$. We pick an ordering of the vertices and also
order the set of pairs of vertices which are connected by at least one
edge. Denote the number of self-loops of the vertex $i$ by
$p_i$, with $1\le i\le\vertex$. Denote the number of edges connecting
the pair $j$ of vertices by $q_j$, with $1\le j\le t$, where
$t$ is the total number of connected pairs of vertices. By
Lemma~\ref{lem:edgesymfac} the edge symmetry factor
of $\gamma$ is given by
$S^\gamma_{\text{edge}}=(\prod_{i=1}^\vertex\, 2^{p_i}\cdot
p_i!)(\prod_{j=1}^t\,q_j!)$ with $\edge=\sum_{i=1}^\vertex p_i+
\sum_{j=1}^t q_j$. Since the graph $\gamma$ has the property that each
vertex has at least one external edge, its vertices are distinguishable
and it has no non-trivial vertex
symmetries: $S^\gamma_{\text{vertex}}=1$ and
$S_{\text{edge}}^\gamma=S^\gamma$ (as any symmetry is an edge
symmetry). We check from which graphs
with $\edge-1$ internal edges $\gamma$ is
generated by the recursion formula (\ref{eq:recomega}) and how
many times it is generated. It turns out that we can think of each
internal edge of $\gamma$ as contributing with a factor of
$1/(\edge\cdot S^\gamma)$ as follows:

(i) Consider the   $i^{\mbox{\tiny{th}}}$ vertex of $\gamma$
endowed with $p_i$ self-loops. Shrinking one of these self-loops
yields a graph $\gamma'$ whose $i^{\mbox{\tiny{th}}}$ vertex has
$p_i-1$ self-loops. Consequently, by Lemma~\ref{lem:edgesymfac}, the
symmetry factor of $\gamma'$ is related to that of $\gamma$ via
$S^{\gamma'}=S^\gamma/(2 p_i)$.  By assumption, the
graph $\gamma'$ corresponds to a term in
$\Dopnew^{\lp-1, \vertex}(\phi(x_1)\cdots\phi(x_n))$ which occurs with
weight given by the inverse of its symmetry factor:
$1/{S^{\gamma'}}= 2 p_i/S^\gamma$. Applying the map
$\Top_i$, which carries the factor $1/2$, to the vertex $i$ of
$\gamma'$ produces the graph $\gamma$ from the graph $\gamma'$ exactly
with factor $p_i/S^\gamma$. Thus, the contribution to
(\ref{eq:recomega}) is $p_i/(\edge\cdot S^\gamma)$. Distributing
this factor between the $p_i$ edges considered yields
$1/(\edge\cdot S^\gamma)$ for each edge considered.

(ii) Consider the $j^{\mbox{\tiny{th}}}$ pair of vertices of
$\gamma$ connected by $q_j$ edges. We assume now the indices of the
vertices forming this pair to be consecutive, given by $k$ and
$k+1$.\footnote{Note that this merely amounts to a particular vertex ordering
of the graph $\gamma$. Since $\gamma$ is a priori
unordered this does not imply any loss of generality.}
Shrinking one of the edges and
fusing the vertices it connects yields a graph $\gamma''$ whose fused
vertex has $r\defeq p_k+p_{k+1}+q_j-1$ self-loops.  Consequently, by
Lemma~\ref{lem:edgesymfac}, the symmetry factor  of $\gamma''$ is
related to that of $\gamma$ as follows:
\[
\frac{1}{2^r}\frac{1}{r!}S^{\gamma''}
=\frac{1}{2^{p_k}p_k!}\frac{1}{2^{p_{k+1}}p_{k+1}!}\frac{1}{q_j!}
S^\gamma .
\]
By assumption, the graph $\gamma''$ corresponds to a term
in $\Dopnew^{\lp, \vertex-1}(\phi(x_1)\cdots\phi(x_n))$ which occurs
with weight given by the inverse of its symmetry factor, i.e.,
\begin{equation}
\frac{1}{S^{\gamma''}}
=\frac{p_k!\,p_{k+1}!\,q_j!}{r!}\cdot\frac{1}{2^{q_j-1}S^\gamma} .
\label{eq:symweight}
\end{equation}
The map $\Qop_k$, when applied to the fused vertex, produces a pair of
vertices occupying the positions $k$ and $k+1$, distributes the $2r$
endings of edges between the two vertices in all possible ways  and
attaches them together by an edge. The action of $\Qop_k$ on the fused
vertex $k$ (leaving out external edges) reads explicitly as 
\begin{align}\label{eq:QRjj1}
\Qop_k\Rop_{k,k}^r & 
 =
 \frac{1}{2}\Rop_{k,k+1}(\Rop_{k,k}+2\Rop_{k,k+1}+\Rop_{k+1,k+1})^r\\
 \label{eq:QRjj2}
& =  \sum_{a=0}^r\sum_{b=0}^a \left( \begin{array}{c}
r\\
a \end{array} \right)\left( \begin{array}{c}
a\\
b \end{array} \right)
2^{a-b-1}\Rop_{k,k}^{r-a}\Rop_{k,k+1}^{a-b+1}\Rop_{k+1,k+1}^b\,.
\end{align}   
Taking into account the external edges,
there are two  terms in equation (\ref{eq:QRjj2}) corresponding to the
graph $\gamma$:  one with   $r-a=p_k$ and $b=p_{k+1}$ and one with
$r-a=p_{k+1}$ and $b=p_k$. The sum of the coefficients of these two
contributions is
\begin{equation}
 2^{q_j-1}\,\frac{r!}{p_k!\,p_{k+1}!\,(q_j-1)!}.
\label{eq:combweight}
\end{equation}
Multiplying (\ref{eq:symweight}) with (\ref{eq:combweight}), we see
that $\Qop_j$ produces $\gamma$ from $\gamma''$ exactly with a factor
$q_j/S^\gamma$ and the contribution to
(\ref{eq:recomega}) is $q_j/(\edge\cdot S^\gamma)$. In other
words, we get a factor of $1/(\edge\cdot S^\gamma)$ for each of
the $q_j$ edges considered.

Since each of the $e=\lp+\vertex-1$ internal edges contributes with a
factor of
$1/(\edge\cdot S^\gamma)$ to the weight of the graph $\gamma$,
the overall contribution is exactly $1/S^\gamma$. This
completes the proof.
\end{proof}

To complete the proof of  Theorem~\ref{thm:main}, we show that the
term in $\Dopnew^{\lp, \vertex}(\phi(x_1)\cdots\phi(x_n))$
corresponding to a connected graph $\gamma$  with  $\lp$ loops,
$\vertex$ vertices and 
external edges whose endpoints are labeled by $x_1,\dots,x_n$, has
weight given by $1/S^\gamma$. If $\gamma$ has external edges
attached to every one of its vertices we
simply recall Lemma~\ref{lem:weightextedges}. Thus, we may now assume
that $\gamma$ has $m$ vertices to which no external leg is
attached. Consider a graph
$\gamma'$ which is constructed from $\gamma$ by attaching an external
edge to every vertex without external edges, choosing arbitrary but
fixed labels $y_1,\dots,y_m$ for the endpoints of external edges in
the process.
By Lemma~\ref{lem:weightextedges}, the graph
$\gamma'$ occurs in the
term on the left hand side of equation (\ref{eq:facomega2}) with
weight $1/S^{\gamma'}$. By Lemma~\ref{lem:allgraphs},
the graph $\gamma$ occurs in the
first factor on the right hand side with some non-zero weight, say
$\alpha$.
Every summand of $\cop^{\vertex-1}(\phi(y_1)\cdots\phi(y_m))$ (recall
formula (\ref{eq:copk}))
which places the endpoints of external edges at the designated
vertices of $\gamma$
to produce $\gamma'$ contributes to the weight of $\gamma'$ in terms
of that of $\gamma$. Any different ways this can happen define a vertex
symmetry of $\gamma$. Furthermore, $\gamma$ can have no more than
these vertex symmetries, since its vertices that already carry external edges
are distinguishable and thus held fixed under any symmetry.
Therefore, using Lemma~\ref{lem:omegafac}
we obtain the formula $1/S^{\gamma'}=\alpha\cdot
S^\gamma_{\text{vertex}}$ by extracting the weights from the
corresponding terms in equation (\ref{eq:facomega2}). Moreover,
$S^{\gamma'}=S^{\gamma'}_{\text{edge}}=S^{\gamma}_{\text{edge}}$.
Thus, using $S^\gamma=S^{\gamma}_{\text{vertex}}\cdot
S^{\gamma}_{\text{edge}}\,$ (Lemma~\ref{lem:sym}), we find
$\alpha=1/S^\gamma$. This completes the proof.

\section{Further recursion relations}
\label{sec:recrel}

Generalizing the case with trees (Section~V in \cite{MeOe:npoint}) we
present an alternative recursion relation satisfied by
$\Dopnew^{l,v}$. This has the advantage over (\ref{eq:recomega})
that it may be translated directly into a recursion relation of the
resulting $n$-point functions $\conn^{l,v}$, related via
Corollary~\ref{cor:loopexpansion}.

\begin{prop}
\label{prop:alternativenew}
Let $v\ge 1$ and $l\ge 0$, but not $v=1$ and $l=0$. Then,
\begin{equation*}
\Dopnew^{l,v}
=\frac{1}{l+v-1}\left(
\Dopnew^{l-1,v}\circ\Top
+ \sum_{a=0}^{l}\sum_{b=1}^{v-1}
\left(\Dopnew^{a,b}\tens\Dopnew^{l-a,v-b}\right)\circ\Qop \right) .
\end{equation*}
It is understood that the first summand does not contribute if $l=0$
while the second does not contribute if $v=1$.
\end{prop}

Before proceeding with the proof we note that this formula has a
straightforward interpretation in terms of sums over weighted
graphs following the correspondence of Section~\ref{sec:grep}.
Namely, the formula states that the weighted sum over graphs
with $l$ loops and  $v$ vertices is given by a sum of two
terms divided by the edge number $e=l+v-1$.
The first term is the sum over all weighted graphs with
$l-1$ loops and  $v$ vertices which have an extra edge
attached, its endpoints being connected to vertices in all possible
ways.
The second term is a sum over
all ordered pairs of weighted graphs with total number of
vertices equal to $v$ and total number of loops equal to
$l$, connected in all possible ways with an edge.

\begin{proof}
The proof proceeds by induction on the number of edges $e=l+v-1$
(recall Lemma \ref{lem:loopnumber}). It is straightforward to check
its validity for $e=1$ by reducing the cases $l=0$, $v=2$ and
$l=1$, $v=1$ to (\ref{eq:recomega}), remembering that $\Dopnew^{0,1}$
is the identity.

We now assume the formula to hold for any edge number
smaller than a fixed $e\ge 2$.  Then for loop number $l$ and vertex number
$v$ such that $e=l+v-1$ we use (\ref{eq:recomega}) to show the
following equality and hence complete the proof.
\begin{flalign*}
& \Dopnew^{l,v} = \frac{1}{l+v-1}\left(\sum_{j=1}^{v-1}\Qop_j
\circ\Dopnew^{l,v-1}
+\sum_{j=1}^{v}\Top_j\circ\Dopnew^{l-1,v}\right)\\
& = \frac{1}{(l+v-1)(l+v-2)}\cdot\\
& \quad \biggl(\sum_{j=1}^{v-1}\Qop_j \circ
\biggl(
 \Dopnew^{l-1,v-1}\circ\Top
+ \sum_{a=0}^{l}\sum_{b=1}^{v-2}
 \left(\Dopnew^{a,b}\tens\Dopnew^{l-a,v-1-b}\right)
\circ\Qop \biggr) \\
& \quad +
\sum_{j=1}^{v}\Top_j\circ\biggl(
 \Dopnew^{l-2,v}\circ\Top
+ \sum_{a=0}^{l-1}\sum_{b=1}^{v-1}
 \left(\Dopnew^{a,b}\tens\Dopnew^{l-1-a,v-b}\right) \circ\Qop
\biggr)\biggr)\\
& = \frac{1}{(l+v-1)(l+v-2)}\cdot\biggl(
 \sum_{j=1}^{v-1}\Qop_j\circ \Dopnew^{l-1,v-1}\circ\Top
+ \sum_{j=1}^{v}\Top_j\circ\Dopnew^{l-2,v}\circ\Top\\
& \quad +
 \sum_{a=0}^{l}\sum_{b=1}^{v-2}\biggl(
 \sum_{j=1}^{b}\Qop_j\circ\Dopnew^{a,b}
 \tens\Dopnew^{l-a,v-1-b}\\
& \quad +
 \sum_{j=1}^{v-1-b}
 \Dopnew^{a,b}\tens
 \Qop_j\circ\Dopnew^{l-a,v-1-b}\biggr)
 \circ\Qop\\
& \quad +
\sum_{a=0}^{l-1}\sum_{b=1}^{v-1}\biggl(\sum_{j=1}^{b}\Top_j\circ
\Dopnew^{a,b}\tens\Dopnew^{l-1-a,v-b}\\
& \quad + \sum_{j=1}^{v-b}
\Dopnew^{a,b}\tens\Top_j\circ\Dopnew^{l-1-a,v-b}\biggr)\circ\Qop
\biggr)\\
& = \frac{1}{(l+v-1)(l+v-2)}\cdot\biggl(
\biggl(\sum_{j=1}^{v-1}\Qop_j\circ
\Dopnew^{l-1,v-1}+\sum_{j=1}^{v}\Top_j\circ\Dopnew^{l-2,v}\biggr)\circ\Top
\\
& \quad +
 \biggl(\sum_{b=1}^{v-2}\biggl(\sum_{j=1}^{b}
 \left(\Qop_j\circ\Dopnew^{0,b}\right) \tens\Dopnew^{l,v-1-b}
+\sum_{j=1}^{v-1-b}\Dopnew^{l,b}\tens
 \left(\Qop_j\circ\Dopnew^{0,v-1-b}\right)\biggr) \\
& \quad + \sum_{a=0}^{l-1}\biggl(\left(\Top\circ
\Dopnew^{a,1}\right)\tens\Dopnew^{l-1-a,v-1}
+ \Dopnew^{a,v-1}\tens\left(\Top\circ\Dopnew^{l-1-a,1}\right)\biggr) \\
&\quad +
\sum_{a=1}^{l}\sum_{b=2}^{v-1}\biggl(
 \sum_{j=1}^{b-1}\Qop_j\circ\Dopnew^{a,b-1}
+ \sum_{j=1}^{b}\Top_j\circ\Dopnew^{a-1,b}\biggr)
 \tens\Dopnew^{l-a,v-b} \\
& \quad +
 \sum_{a=0}^{l-1}\sum_{b=1}^{v-2}
 \Dopnew^{a,b}\tens\biggl(\sum_{j=1}^{v-1-b}\Qop_j
 \circ\Dopnew^{l-a,v-1-b}
 + \sum_{j=1}^{v-b}\Top_j\circ\Dopnew^{l-1-a,v-b}\biggr)\biggr)
 \circ\Qop\biggr)\\
& = \frac{1}{(l+v-1)(l+v-2)}\cdot\biggl(
(l+v-2)\,\Dopnew^{l-1,v}\circ\Top\\
& \quad +
 \biggl(\sum_{b=2}^{v-1}(b-1)\,\Dopnew^{0,b}
 \tens\Dopnew^{l,v-b}
 +\sum_{b=1}^{v-2}(v-b-1)\,\Dopnew^{l,b}\tens\Dopnew^{0,v-b}\\
& \quad + \sum_{a=1}^{l} a\,
 \Dopnew^{a,1}\tens\Dopnew^{l-a,v-1}
 + \sum_{a=0}^{l-1} (l-a)\,
 \Dopnew^{a,v-1}\tens\Dopnew^{l-a,1} \\
&\quad +
 \sum_{a=1}^{l}\sum_{b=2}^{v-1} (a+b-1)\,\Dopnew^{a,b}
 \tens\Dopnew^{l-a,v-b}\\
& \quad +
\sum_{a=0}^{l-1}\sum_{b=1}^{v-2} (l-a+v-b-1)\,\Dopnew^{a,b}\tens
\Dopnew^{l-a,v-b}\biggr)\circ\Qop\biggr)\\
& = \frac{1}{(l+v-1)(l+v-2)}\cdot\biggl(
 (l+v-2)\,\Dopnew^{l-1,v}\circ\Top \\
& \quad +
\biggl(\sum_{a=0}^{l}\sum_{b=1}^{v-1}
(a+b-1)\,\Dopnew^{a,b}\tens\Dopnew^{l-a,v-b} \\
& \quad +
\sum_{a=0}^{l}\sum_{b=1}^{v-1} (l-a+v-b-1)\,\Dopnew^{a,b}\tens
\Dopnew^{l-a,v-b}\biggr)\circ\Qop\biggr)\\
& =\frac{1}{l+v-1}\left(\Dopnew^{l-1,v}\circ\Top
 + \sum_{a=0}^{l}\sum_{b=1}^{v-1}
 \left(\Dopnew^{a,b}\tens\Dopnew^{l-a,v-b}\right)
\circ\Qop\right) .
\end{flalign*}
\end{proof}

Combining this result with Corollary~\ref{cor:loopexpansion} yields
the corresponding recursion equation for $\conn^{l,v}$.
\begin{cor}
\label{cor:recnew}
Let $v\ge 1$ and $l\ge 0$, but not $v=1$ and $l=0$. Then,
\begin{equation*}
\conn^{l,v}
=\frac{1}{l+v-1}\left(
\conn^{l-1,v}\circ\Top
+ \sum_{a=0}^{l}\sum_{b=1}^{v-1}
\left(\conn^{a,b}\tens\conn^{l-a,v-b}\right)\circ\Qop \right) .
\end{equation*}
It is understood that the first summand does not contribute if $l=0$
while the second does not contribute if $v=1$.
\end{cor}

\section{Discussion and Conclusion}
\label{sec:discuss}

The results of the present paper can be seen as an extension of those
of \cite{MeOe:npoint}, where only tree graphs were
generated. Accordingly, many points in the discussion of the main
result in that paper
extend to the present setting. In particular, this
applies to the algorithmic aspects and to the inclusion of
fermions. We refer the reader to Sections~VI.C and VI.D of
\cite{MeOe:npoint} for details. Here we shall only touch these points
briefly and highlight differences arising through the inclusion of
graphs with loops.

The generation of the graphs in their algebraic
representation via the recursion formula (\ref{eq:recomega})
has the structure of an algorithm. Indeed, this algorithmic structure
can be used directly and efficiently in implementing concrete
calculations of (loop) graphs. In doing so, external edges may be fixed
from the beginning and $\Dopnew^{\lp,\vertex}$ as applied to the
external edges is calculated recursively rather than as an abstract map.
An important aspect for the
efficiency of concrete calculations is to discard graphs that do not
contribute. In typical quantum field theoretic calculations, the
vertex function is such that only vertices with a minimum valence
(usually three)
contribute. In the case of tree graphs this allows the restriction of
the coproduct implicit in the operator $Q_i$ in the recursion formula
(\ref{eq:recomega}) \cite{MeOe:npoint}. Concretely,
the coproduct (\ref{eq:coproduct})
may be replaced by a \emph{truncated} coproduct $\cop_{\ge k}$ with
$k\ge 1$. This
is defined by removing from the right hand side of
(\ref{eq:coproduct}) all terms where the number of elements in $I_1$
or $I_2$ is smaller than $k$. This will prevent graphs from being
generated who have vertices with valence smaller than $k+1$. If only
tree graphs are considered this is consistent with the recursion
process. More precisely, a graph with all vertices of valence at least
$k+1$ cannot be generated by $Q_i$ from a graph with at least one vertex
having valence smaller than $k+1$. The analogous statement is
not true for the operator $T_i$. Hence, considering loop graphs as
well (recall that $T_i$ increases loop number), we can no longer
globally restrict the coproduct. However, if we are interested in
graphs only up to a maximal loop number $m$, we may still restrict the
coproduct in $Q_i$ in certain instances. These are precisely the
instances when a later application of $T_i$ to a graph cannot occur,
i.e., when the graph has already the maximal loop number $m$.

The restriction on the valence of vertices to be at least $a$, where
$a\ge 3$, leads to another obvious limit we can impose on the
algorithm. Namely, for a given number of loops $m$ and a given number
of external edges $n$ there is an upper bound $b= (n+2m-2)/(a-2)$
on the number of
vertices a graph can have. Thus, in this case we only need to compute
$\Dopnew^{l,v}$ for $l\le m$ and
$v\le b$.

We now turn to the question of the implementation of fermions. Here
the situation is not at all changed by the extension to loop
graphs. Namely, the whole formalism is completely functorial and
carries over immediately to the case that the vector space $V$ of
field operators is a $\Z_2$-graded space. (Recall that this means that
$V$ is a direct sum of a bosonic and fermionic part.) Concretely,
certain field operators will anticommute which introduces minus signs
in front the summands in (\ref{eq:coproduct}) and (\ref{eq:copk})
which correspond to odd permutations of such field operators. In
contrast, all formulas appearing in Section~\ref{sec:loop} and
\ref{sec:recrel} remain
unchanged as the $\Z_2$-grading is completely implicit there.

The algorithm to generate tree graphs was applied in two contexts in
\cite{MeOe:npoint}: To relate connected $n$-point functions with
1-particle irreducible ones and to generate all tree graphs using the
vertex functions. In both cases renormalization does not introduce any
alteration. This is different in the present situation where we
interpret the algorithms of Section~\ref{sec:loop} and
\ref{sec:recrel} as generating all
connected graphs using the vertex functions. Renormalization, via
counter-terms, alters this process considerably. Thus, it would be
highly desirable to include the renormalization process into the present
framework. At this point we have very little to say about this, except
to point out that the algorithms of Section~\ref{sec:loop} and
\ref{sec:recrel} are
naturally organized as a recursion by loop order, which might
facilitate the task.

\subsection*{Acknowledgments}

We would like to thank Christian Brouder who, after learning about our
main result, suggested the recursion formula of
Proposition~\ref{prop:alternativenew}, which we subsequently
proved. One of the authors (\^A.~M.) was supported through a
fellowship provided by Funda\c c\~ao Calouste Gulbenkian 65709.

\appendix
\section{Appendix}

This appendix shows all graphs without external edges and
with up to three edges computed as contributions to
$\Dopnew^{\lp,\vertex}(\one)$
via (\ref{eq:recomega}).  The factors in front of the graphs are the
inverses of their
symmetry factors of Definition~\ref{def:symfac}, see
Theorem~\ref{thm:main}.
\vspace{1cm}\\
\setlength{\unitlength}{0.00083333in}
\begingroup\makeatletter\ifx\SetFigFont\undefined%
\gdef\SetFigFont#1#2#3#4#5{%
  \reset@font\fontsize{#1}{#2pt}%
  \fontfamily{#3}\fontseries{#4}\fontshape{#5}%
  \selectfont}%
\fi\endgroup%
{\renewcommand{\dashlinestretch}{30}
\begin{picture}(1143,264)(0,-10)
\texture{44555555 55aaaaaa aa555555 55aaaaaa aa555555 55aaaaaa aa555555 55aaaaaa 
	aa555555 55aaaaaa aa555555 55aaaaaa aa555555 55aaaaaa aa555555 55aaaaaa 
	aa555555 55aaaaaa aa555555 55aaaaaa aa555555 55aaaaaa aa555555 55aaaaaa 
	aa555555 55aaaaaa aa555555 55aaaaaa aa555555 55aaaaaa aa555555 55aaaaaa }
\put(1060,166){\shade\ellipse{150}{150}}
\put(1060,166){\ellipse{150}{150}}
\put(0,58){\makebox(0,0)[lb]{{\SetFigFont{12}{14.4}{\rmdefault}{\mddefault}{\updefault}$\Dopnew^{0,1}(\one)\,\,=$}}}
\end{picture}
}
\vspace{1cm}\\
\setlength{\unitlength}{0.00083333in}
\begingroup\makeatletter\ifx\SetFigFont\undefined%
\gdef\SetFigFont#1#2#3#4#5{%
  \reset@font\fontsize{#1}{#2pt}%
  \fontfamily{#3}\fontseries{#4}\fontshape{#5}%
  \selectfont}%
\fi\endgroup%
{\renewcommand{\dashlinestretch}{30}
\begin{picture}(1630,274)(0,-10)
\texture{44555555 55aaaaaa aa555555 55aaaaaa aa555555 55aaaaaa aa555555 55aaaaaa 
	aa555555 55aaaaaa aa555555 55aaaaaa aa555555 55aaaaaa aa555555 55aaaaaa 
	aa555555 55aaaaaa aa555555 55aaaaaa aa555555 55aaaaaa aa555555 55aaaaaa 
	aa555555 55aaaaaa aa555555 55aaaaaa aa555555 55aaaaaa aa555555 55aaaaaa }
\path(1262,168)(1562,168)
\path(1262,168)(1562,168)
\put(1247,176){\shade\ellipse{150}{150}}
\put(1247,176){\ellipse{150}{150}}
\put(1547,176){\shade\ellipse{150}{150}}
\put(1547,176){\ellipse{150}{150}}
\put(0,58){\makebox(0,0)[lb]{{\SetFigFont{12}{14.4}{\rmdefault}{\mddefault}{\updefault}$\Dopnew^{0,2}(\one)\,\,=\,\,\frac{1}{2}$}}}
\end{picture}
}
\vspace{1cm}\\
\setlength{\unitlength}{0.00083333in}
\begingroup\makeatletter\ifx\SetFigFont\undefined%
\gdef\SetFigFont#1#2#3#4#5{%
  \reset@font\fontsize{#1}{#2pt}%
  \fontfamily{#3}\fontseries{#4}\fontshape{#5}%
  \selectfont}%
\fi\endgroup%
{\renewcommand{\dashlinestretch}{30}
\begin{picture}(1331,449)(0,-10)
\put(1248,305){\ellipse{120}{246}}
\texture{44555555 55aaaaaa aa555555 55aaaaaa aa555555 55aaaaaa aa555555 55aaaaaa 
	aa555555 55aaaaaa aa555555 55aaaaaa aa555555 55aaaaaa aa555555 55aaaaaa 
	aa555555 55aaaaaa aa555555 55aaaaaa aa555555 55aaaaaa aa555555 55aaaaaa 
	aa555555 55aaaaaa aa555555 55aaaaaa aa555555 55aaaaaa aa555555 55aaaaaa }
\put(1248,178){\shade\ellipse{150}{150}}
\put(1248,178){\ellipse{150}{150}}
\put(0,58){\makebox(0,0)[lb]{{\SetFigFont{12}{14.4}{\rmdefault}{\mddefault}{\updefault}$\Dopnew^{1,1}(\one)\,\,=\,\,\frac{1}{2}$}}}
\end{picture}
}
\vspace{1cm}\\
\setlength{\unitlength}{0.00083333in}
\begingroup\makeatletter\ifx\SetFigFont\undefined%
\gdef\SetFigFont#1#2#3#4#5{%
  \reset@font\fontsize{#1}{#2pt}%
  \fontfamily{#3}\fontseries{#4}\fontshape{#5}%
  \selectfont}%
\fi\endgroup%
{\renewcommand{\dashlinestretch}{30}
\begin{picture}(1925,279)(0,-10)
\texture{44555555 55aaaaaa aa555555 55aaaaaa aa555555 55aaaaaa aa555555 55aaaaaa 
	aa555555 55aaaaaa aa555555 55aaaaaa aa555555 55aaaaaa aa555555 55aaaaaa 
	aa555555 55aaaaaa aa555555 55aaaaaa aa555555 55aaaaaa aa555555 55aaaaaa 
	aa555555 55aaaaaa aa555555 55aaaaaa aa555555 55aaaaaa aa555555 55aaaaaa }
\path(1242,181)(1542,181)
\path(1242,181)(1542,181)
\path(1542,181)(1842,181)
\path(1542,181)(1842,181)
\put(1242,181){\shade\ellipse{150}{150}}
\put(1242,181){\ellipse{150}{150}}
\put(1542,181){\shade\ellipse{150}{150}}
\put(1542,181){\ellipse{150}{150}}
\put(1842,181){\shade\ellipse{150}{150}}
\put(1842,181){\ellipse{150}{150}}
\put(0,58){\makebox(0,0)[lb]{{\SetFigFont{12}{14.4}{\rmdefault}{\mddefault}{\updefault}$\Dopnew^{0,3}(\one)\,\,=\,\,\frac{1}{2}$}}}
\end{picture}
}
\vspace{1cm}\\
\setlength{\unitlength}{0.00083333in}
\begingroup\makeatletter\ifx\SetFigFont\undefined%
\gdef\SetFigFont#1#2#3#4#5{%
  \reset@font\fontsize{#1}{#2pt}%
  \fontfamily{#3}\fontseries{#4}\fontshape{#5}%
  \selectfont}%
\fi\endgroup%
{\renewcommand{\dashlinestretch}{30}
\begin{picture}(2862,294)(0,-10)
\put(1300,186){\ellipse{240}{114}}
\put(2630,196){\ellipse{300}{150}}
\path(1397,186)(1721,186)(1709,186)
\texture{44555555 55aaaaaa aa555555 55aaaaaa aa555555 55aaaaaa aa555555 55aaaaaa 
	aa555555 55aaaaaa aa555555 55aaaaaa aa555555 55aaaaaa aa555555 55aaaaaa 
	aa555555 55aaaaaa aa555555 55aaaaaa aa555555 55aaaaaa aa555555 55aaaaaa 
	aa555555 55aaaaaa aa555555 55aaaaaa aa555555 55aaaaaa aa555555 55aaaaaa }
\put(2779,194){\shade\ellipse{150}{150}}
\put(2779,194){\ellipse{150}{150}}
\put(2474,193){\shade\ellipse{150}{150}}
\put(2474,193){\ellipse{150}{150}}
\put(1721,189){\shade\ellipse{150}{150}}
\put(1721,189){\ellipse{150}{150}}
\put(1411,185){\shade\ellipse{150}{150}}
\put(1411,185){\ellipse{150}{150}}
\put(1920,58){\makebox(0,0)[lb]{{\SetFigFont{12}{14.4}{\rmdefault}{\mddefault}{\updefault}$+\,\,\frac{1}{2^ 2}$}}}
\put(0,58){\makebox(0,0)[lb]{{\SetFigFont{12}{14.4}{\rmdefault}{\mddefault}{\updefault}$\Dopnew^{1,2}(\one)\,\,=\,\,\frac{1}{2}$}}}
\end{picture}
}
\vspace{1cm}\\
\setlength{\unitlength}{0.00083333in}
\begingroup\makeatletter\ifx\SetFigFont\undefined%
\gdef\SetFigFont#1#2#3#4#5{%
  \reset@font\fontsize{#1}{#2pt}%
  \fontfamily{#3}\fontseries{#4}\fontshape{#5}%
  \selectfont}%
\fi\endgroup%
{\renewcommand{\dashlinestretch}{30}
\begin{picture}(1396,528)(0,-10)
\put(1313,384){\ellipse{120}{246}}
\put(1313,129){\ellipse{120}{246}}
\texture{44555555 55aaaaaa aa555555 55aaaaaa aa555555 55aaaaaa aa555555 55aaaaaa 
	aa555555 55aaaaaa aa555555 55aaaaaa aa555555 55aaaaaa aa555555 55aaaaaa 
	aa555555 55aaaaaa aa555555 55aaaaaa aa555555 55aaaaaa aa555555 55aaaaaa 
	aa555555 55aaaaaa aa555555 55aaaaaa aa555555 55aaaaaa aa555555 55aaaaaa }
\put(1313,257){\shade\ellipse{150}{150}}
\put(1313,257){\ellipse{150}{150}}
\put(0,137){\makebox(0,0)[lb]{{\SetFigFont{12}{14.4}{\rmdefault}{\mddefault}{\updefault}$\Dopnew^{2,1}(\one)\,\,=\,\,\frac{1}{2^3}$}}}
\end{picture}
}
\vspace{1cm}\\
\setlength{\unitlength}{0.00083333in}
\begingroup\makeatletter\ifx\SetFigFont\undefined%
\gdef\SetFigFont#1#2#3#4#5{%
  \reset@font\fontsize{#1}{#2pt}%
  \fontfamily{#3}\fontseries{#4}\fontshape{#5}%
  \selectfont}%
\fi\endgroup%
{\renewcommand{\dashlinestretch}{30}
\begin{picture}(3413,701)(0,-10)
\texture{44555555 55aaaaaa aa555555 55aaaaaa aa555555 55aaaaaa aa555555 55aaaaaa 
	aa555555 55aaaaaa aa555555 55aaaaaa aa555555 55aaaaaa aa555555 55aaaaaa 
	aa555555 55aaaaaa aa555555 55aaaaaa aa555555 55aaaaaa aa555555 55aaaaaa 
	aa555555 55aaaaaa aa555555 55aaaaaa aa555555 55aaaaaa aa555555 55aaaaaa }
\path(3330,343)(3030,343)
\path(3330,343)(3030,343)
\path(2880,603)(3040,333)
\path(2890,93)(3050,363)
\path(1267,335)(1567,335)
\path(1267,335)(1567,335)
\path(1567,335)(1867,335)
\path(1567,335)(1867,335)
\path(1867,335)(2167,335)
\path(1867,335)(2167,335)
\put(3030,343){\shade\ellipse{150}{150}}
\put(3030,343){\ellipse{150}{150}}
\put(3330,343){\shade\ellipse{150}{150}}
\put(3330,343){\ellipse{150}{150}}
\put(2885,603){\shade\ellipse{150}{150}}
\put(2885,603){\ellipse{150}{150}}
\put(2885,83){\shade\ellipse{150}{150}}
\put(2885,83){\ellipse{150}{150}}
\put(1537,328){\shade\ellipse{150}{150}}
\put(1537,328){\ellipse{150}{150}}
\put(1837,328){\shade\ellipse{150}{150}}
\put(1837,328){\ellipse{150}{150}}
\put(2137,328){\shade\ellipse{150}{150}}
\put(2137,328){\ellipse{150}{150}}
\put(1237,336){\shade\ellipse{150}{150}}
\put(1237,336){\ellipse{150}{150}}
\put(2310,223){\makebox(0,0)[lb]{{\SetFigFont{12}{14.4}{\rmdefault}{\mddefault}{\updefault}$+\,\,\frac{1}{3!}$}}}
\put(0,223){\makebox(0,0)[lb]{{\SetFigFont{12}{14.4}{\rmdefault}{\mddefault}{\updefault}$\Dopnew^{0,4}(\one)\,\,=\,\,\frac{1}{2}$}}}
\end{picture}
}
\vspace{1cm}\\
\setlength{\unitlength}{0.00083333in}
\begingroup\makeatletter\ifx\SetFigFont\undefined%
\gdef\SetFigFont#1#2#3#4#5{%
  \reset@font\fontsize{#1}{#2pt}%
  \fontfamily{#3}\fontseries{#4}\fontshape{#5}%
  \selectfont}%
\fi\endgroup%
{\renewcommand{\dashlinestretch}{30}
\begin{picture}(5792,529)(0,-10)
\put(2627,307){\ellipse{120}{246}}
\put(3658,180){\ellipse{240}{114}}
\path(2269,183)(2629,183)(2635,183)
\path(2569,183)(2929,183)(2935,183)
\path(4460,187)(4100,187)(4094,187)
\path(4160,187)(3800,187)(3794,187)
\put(5276,185){\ellipse{300}{150}}
\path(5385,182)(5745,182)(5751,182)
\path(1434,425)(1291,177)(1577,177)(1434,425)
\texture{44555555 55aaaaaa aa555555 55aaaaaa aa555555 55aaaaaa aa555555 55aaaaaa 
	aa555555 55aaaaaa aa555555 55aaaaaa aa555555 55aaaaaa aa555555 55aaaaaa 
	aa555555 55aaaaaa aa555555 55aaaaaa aa555555 55aaaaaa aa555555 55aaaaaa 
	aa555555 55aaaaaa aa555555 55aaaaaa aa555555 55aaaaaa aa555555 55aaaaaa }
\put(2629,188){\shade\ellipse{150}{150}}
\put(2629,188){\ellipse{150}{150}}
\put(2917,194){\shade\ellipse{150}{150}}
\put(2917,194){\ellipse{150}{150}}
\put(2337,189){\shade\ellipse{150}{150}}
\put(2337,189){\ellipse{150}{150}}
\put(1431,433){\shade\ellipse{150}{150}}
\put(1431,433){\ellipse{150}{150}}
\put(1284,177){\shade\ellipse{150}{150}}
\put(1284,177){\ellipse{150}{150}}
\put(1587,184){\shade\ellipse{150}{150}}
\put(1587,184){\ellipse{150}{150}}
\put(4100,182){\shade\ellipse{150}{150}}
\put(4100,182){\ellipse{150}{150}}
\put(4392,181){\shade\ellipse{150}{150}}
\put(4392,181){\ellipse{150}{150}}
\put(3794,185){\shade\ellipse{150}{150}}
\put(3794,185){\ellipse{150}{150}}
\put(5127,185){\shade\ellipse{150}{150}}
\put(5127,185){\ellipse{150}{150}}
\put(5421,185){\shade\ellipse{150}{150}}
\put(5421,185){\ellipse{150}{150}}
\put(5709,191){\shade\ellipse{150}{150}}
\put(5709,191){\ellipse{150}{150}}
\put(1790,63){\makebox(0,0)[lb]{{\SetFigFont{12}{14.4}{\rmdefault}{\mddefault}{\updefault}$+\,\,\frac{1}{2^2}$}}}
\put(0,58){\makebox(0,0)[lb]{{\SetFigFont{12}{14.4}{\rmdefault}{\mddefault}{\updefault}$\Dopnew^{1,3}(\one)\,\,=\,\,\frac{1}{3!}$}}}
\put(3120,63){\makebox(0,0)[lb]{{\SetFigFont{12}{14.4}{\rmdefault}{\mddefault}{\updefault}$+\,\,\frac{1}{2}$}}}
\put(4610,68){\makebox(0,0)[lb]{{\SetFigFont{12}{14.4}{\rmdefault}{\mddefault}{\updefault}$+\,\,\frac{1}{2}$}}}
\end{picture}
}
\vspace{1cm}\\
\setlength{\unitlength}{0.00083333in}
\begingroup\makeatletter\ifx\SetFigFont\undefined%
\gdef\SetFigFont#1#2#3#4#5{%
  \reset@font\fontsize{#1}{#2pt}%
  \fontfamily{#3}\fontseries{#4}\fontshape{#5}%
  \selectfont}%
\fi\endgroup%
{\renewcommand{\dashlinestretch}{30}
\begin{picture}(5516,507)(0,-10)
\put(1356,363){\ellipse{120}{246}}
\put(1356,129){\ellipse{120}{246}}
\put(3020,252){\ellipse{240}{114}}
\put(2483,248){\ellipse{240}{114}}
\put(3852,243){\ellipse{240}{114}}
\path(1338,244)(1662,244)(1650,244)
\path(2923,252)(2599,252)(2611,252)
\put(5287,247){\ellipse{300}{150}}
\path(5109,244)(5433,244)(5421,244)
\put(4122,245){\ellipse{300}{150}}
\texture{44555555 55aaaaaa aa555555 55aaaaaa aa555555 55aaaaaa aa555555 55aaaaaa 
	aa555555 55aaaaaa aa555555 55aaaaaa aa555555 55aaaaaa aa555555 55aaaaaa 
	aa555555 55aaaaaa aa555555 55aaaaaa aa555555 55aaaaaa aa555555 55aaaaaa 
	aa555555 55aaaaaa aa555555 55aaaaaa aa555555 55aaaaaa aa555555 55aaaaaa }
\put(1662,247){\shade\ellipse{150}{150}}
\put(1662,247){\ellipse{150}{150}}
\put(1352,243){\shade\ellipse{150}{150}}
\put(1352,243){\ellipse{150}{150}}
\put(2599,249){\shade\ellipse{150}{150}}
\put(2599,249){\ellipse{150}{150}}
\put(2909,253){\shade\ellipse{150}{150}}
\put(2909,253){\ellipse{150}{150}}
\put(5433,247){\shade\ellipse{150}{150}}
\put(5433,247){\ellipse{150}{150}}
\put(5139,247){\shade\ellipse{150}{150}}
\put(5139,247){\ellipse{150}{150}}
\put(4270,245){\shade\ellipse{150}{150}}
\put(4270,245){\ellipse{150}{150}}
\put(3961,241){\shade\ellipse{150}{150}}
\put(3961,241){\ellipse{150}{150}}
\put(3250,121){\makebox(0,0)[lb]{{\SetFigFont{12}{14.4}{\rmdefault}{\mddefault}{\updefault}$+\,\,\frac{1}{2^ 2}$}}}
\put(4490,121){\makebox(0,0)[lb]{{\SetFigFont{12}{14.4}{\rmdefault}{\mddefault}{\updefault}$+\,\,\frac{1}{2\cdot3!}$}}}
\put(1865,116){\makebox(0,0)[lb]{{\SetFigFont{12}{14.4}{\rmdefault}{\mddefault}{\updefault}$+\,\,\frac{1}{2^ 3}$}}}
\put(0,116){\makebox(0,0)[lb]{{\SetFigFont{12}{14.4}{\rmdefault}{\mddefault}{\updefault}$\Dopnew^{2,2}(\one)\,\,=\,\,\frac{1}{2^3}$}}}
\end{picture}
}
\vspace{1cm}\\
\setlength{\unitlength}{0.00083333in}
\begingroup\makeatletter\ifx\SetFigFont\undefined%
\gdef\SetFigFont#1#2#3#4#5{%
  \reset@font\fontsize{#1}{#2pt}%
  \fontfamily{#3}\fontseries{#4}\fontshape{#5}%
  \selectfont}%
\fi\endgroup%
{\renewcommand{\dashlinestretch}{30}
\begin{picture}(1903,450)(0,-10)
\put(1650,305){\ellipse{120}{246}}
\put(1535,176){\ellipse{240}{114}}
\put(1775,176){\ellipse{240}{114}}
\texture{44555555 55aaaaaa aa555555 55aaaaaa aa555555 55aaaaaa aa555555 55aaaaaa 
	aa555555 55aaaaaa aa555555 55aaaaaa aa555555 55aaaaaa aa555555 55aaaaaa 
	aa555555 55aaaaaa aa555555 55aaaaaa aa555555 55aaaaaa aa555555 55aaaaaa 
	aa555555 55aaaaaa aa555555 55aaaaaa aa555555 55aaaaaa aa555555 55aaaaaa }
\put(1655,173){\shade\ellipse{150}{150}}
\put(1655,173){\ellipse{150}{150}}
\put(0,58){\makebox(0,0)[lb]{{\SetFigFont{12}{14.4}{\rmdefault}{\mddefault}{\updefault}$\Dopnew^{3,1}(\one)\,\,=\,\,\frac{1}{2^ 3\cdot3!}$}}}
\end{picture}
}
\\

\bibliographystyle{amsordx}
\bibliography{stdrefs}

\end{document}